\documentclass[showpacs,twocolumn,nofootinbib]{revtex4}
\usepackage{amsfonts}
\usepackage{graphicx}
\usepackage{amsmath}
\usepackage{amssymb}
\usepackage{amsthm}

\begin{document}

\title{Soft singularity crossing and transformation of matter properties}
\author{Zolt\'{a}n Keresztes$^{1,2}$, L\'{a}szl\'{o} \'{A}. Gergely$^{1,2}$,
Alexander Yu. Kamenshchik$^{3,4}$, Vittorio Gorini$^{5}$, David Polarski$%
^{6} $}
\affiliation{$^{1}$ Department of Theoretical Physics, University of Szeged, Tisza Lajos
krt 84-86, Szeged 6720, Hungary \\
$^{2}$ Department of Experimental Physics, University of Szeged, D\'{o}m T 
\'{e}r 9, Szeged 6720, Hungary\\
$^{3}$ Dipartimento di Fisica e Astronomia, Universit\`a di Bologna and
INFN, via Irnerio 46, 40126 Bologna, Italy \\
$^{4}$ L.D. Landau Institute for Theoretical Physics, Russian Academy of
Sciences, Kosygin street 2, 119334 Moscow, Russia\\
$^{5}~$Dipartimento di Scienza e Alta Tecnologia, Universit\`{a}
dell'Insubria, Via Valleggio 11, 22100 Como, Italy\\
$^{6}~$Laboratoire Charles Coulomb UMR 5221, Universit\'{e} Montpellier 2
and CNRS, F-34095 Montpellier, France}

\begin{abstract}
We investigate particular cosmological models, based either on tachyon
fields or on perfect fluids, for which soft future singularities arise in a
natural way. Our main result is the description of a smooth crossing of the
soft singularity in models with an anti-Chaplygin gas or with a particular
tachyon field in the presence of dust. Such a crossing is made possible by
certain transformations of matter properties. Some of these cosmological
evolutions involving tachyons are compatible with SNIa data. 
We compute numerically their dynamics involving a first soft singularity
crossing, a turning point and a second soft singulatity crossing during
recollapse, ending in a Big Crunch singularity.
\end{abstract}

\maketitle

\section{Introduction}

In a recent paper \cite{Paradox} we have investigated the possibility of
soft singularity crossing in a model where a flat Friedmann universe was
filled with dust and anti-Chaplygin gas. A soft singularity appears in an
expanding universe when the pressure of the anti-Chaplygin gas diverges,
causing $\ddot{a}\rightarrow -\infty $ ($a$ is the scale factor and the dot
denotes the derivative with respect to cosmic time), while $a$ and $\dot{a}$
remain finite. The energy density of the anti-Chaplygin gas vanishes at the
singularity while the energy density of dust remains finite there, thus
implying $\dot{a}>0$. The geodesic equations remain regular at the
singularity and, therefore, they can be continued through. This allows the
universe to cross the soft singularity. Then, a smooth evolution of the
universe would require further expansion. However, in this case the energy
density of the anti-Chaplygin gas would become imaginary and hence ill
defined. This contradiction and the fact that the geodesics can be continued
through leads to a paradox. In \cite{Paradox} we have solved this paradox by
relaxing the smoothness condition, leading to the redefinition of
cosmological quantities as distributions. With this redefinition, it turns
out that the universe can revert abruptly from expansion to contraction.

In the present work we study an alternative possibility for the continuation
of geodesics across the singularity, requiring the continuity of the
spacetime evolution at the expense of certain transformations of matter
properties.

The paper is organized as follows. In Sec. II we discuss the sudden
singularities that arise in a class of flat Friedmann models such as those
describing universes filled with the anti-Chaplygin gas without or with
dust, or driven by a specific tachyon field \cite{tach0}, again without and
with dust. In Sec. III we describe the crossing of a soft singularity with
accompanying transformations of matter in the above-mentioned models.
Section IV presents numerical results of the evolution of universes filled
with a tachyon field and dust, compatible with SNIa data. Concluding remarks
are presented in Sec. V. We choose units $c=1$ and $8\pi G/3=1$.

\section{Sudden singularities in flat Friedmann universes\label%
{FriedmannFlat}}

The line element squared of a flat Friedmann universe can be written as%
\begin{equation}
ds^{2}=dt^{2}-a^{2}\left( t\right) \sum_{\alpha }\left( dx^{\alpha }\right)
^{2},
\end{equation}%
where $x^{\alpha }$ ($\alpha =1,2,3$) are spatial Cartesian coordinates. The
evolution of the universe is governed by the Raychaudhuri (second Friedmann)
equation%
\begin{equation}
\dot{H}=-\frac{3}{2}\left( \rho +p\right) ,  \label{Raych}
\end{equation}%
and by the continuity equation for the fluid, 
\begin{equation}
\dot{\rho}+3H\left( \rho +p\right) =0,  \label{cont}
\end{equation}%
where, as usual, $\rho $ is the energy density, $p$ is the pressure of
matter and $H\equiv \dot{a}/a$ is the Hubble parameter. The first Friedmann
equation is 
\begin{equation}
H^{2}=\rho .  \label{Friedmann}
\end{equation}

Sudden singularities are characterized by finite $H_{S}$ and $\dot{H}%
_{S}=-\infty $ (finite $\dot{a}_{S}$ and $\ddot{a}_{S}=-\infty $) at some
finite scale factor $a_{S}$. Here, the subscript $S$ denotes the respective
quantities evaluated at the singularity. 
These conditions can be formulated in terms of energy density and pressure
of the fluid. The Friedmann (\ref{Friedmann}) and Raychaudhuri (\ref{Raych})
equations show that the total energy density $\rho _{S}$ is nonnegative and
finite while the pressure diverges $p_{S}=\infty $.

It was shown in \cite{Lazkoz}, \cite{tach2} and \cite{Paradox} that the
geodesics can be continued across such sudden singularities as the geodesic
equations are regular there. The singularity is weak (soft) according to the
definitions of both Tipler \cite{Tipler} and Kr\'{o}lak \cite{Krolak}.
Although the tidal forces become infinite, the extended objects are not
necessarily crushed when reaching the singularity.

\subsection{Big brake\label{BB}}

A special case of sudden singularity is the Big Brake singularity, occurring
when the energy density vanishes at the singularity, $\rho _{S}=0$ \cite%
{tach0}.

\subsubsection{Anti-Chaplygin gas}

One of the simplest models where the Big Brake singularity arises is the
anti-Chaplygin gas \cite{tach0}. This is a perfect fluid with the equation
of state%
\begin{equation}
p=\frac{A}{\rho }~,  \label{anti-Chap}
\end{equation}%
where $A>0$, as opposed to the Chaplygin gas \cite{we-Chap,we-Chap1} which
has the equation of state $p = -A/\rho$. The equation of state (\ref%
{anti-Chap}) arises, for example, in the theory of wiggly strings \cite%
{wiggly}.

Applied to the anti-Chaplygin gas, the continuity equation (\ref{cont})
gives the following dependence of the energy density on the scale factor: 
\begin{equation}
\rho _{ACh}=\sqrt{\frac{B}{a^{6}}-A}~,  \label{solrhoach}
\end{equation}%
where $B$ is a positive constant, which determines the initial condition.
When the scale factor approaches the value 
\begin{equation}
a_{S}=\left( \frac{B}{A}\right) ^{\frac{1}{6}},  \label{radius-sing}
\end{equation}%
during the expansion of the universe, the energy density of the
anti-Chaplygin gas vanishes, and its pressure grows to infinity.
Accordingly, the deceleration also becomes infinite.

As it was shown in \cite{Paradox}, after crossing of this singularity the
universe starts contracting towards a Big Crunch.

\subsubsection{The tachyon field with trigonometric potential and transition
to a Born-Infeld type pseudotachyon field}

A Big Brake singularity was first found in a specific tachyon model
introduced in \cite{tach0}. The Lagrangian density of a tachyon field is 
\cite{Sen} 
\begin{equation}
L=-V(T)\sqrt{1-g^{ij}\left( \partial _{i}T\right) \left( \partial
_{j}T\right) },  \label{L}
\end{equation}%
where $V(T)$ is a potential. For a spatially homogeneous field $T(t)$, the
expression (\ref{L}) becomes 
\begin{equation}
L=-V(T)\sqrt{1-g^{tt}\dot{T}^{2}}.  \label{L1}
\end{equation}%
This field corresponds to an ideal fluid with energy density 
\begin{equation}
\rho _{T}=\frac{V(T)}{\sqrt{1-\dot{T}^{2}}},  \label{rho}
\end{equation}%
and pressure 
\begin{equation}
p_{T}=-V(T)\sqrt{1-\dot{T}^{2}}.  \label{p}
\end{equation}%
The Lagrangian density as well as $\rho _{T}$ and $p_{T}$ are well defined
for $\dot{T}^{2}\leq 1$. The field equation is%
\begin{equation}
\frac{\ddot{T}}{1-\dot{T}^{2}}+3H\dot{T}+\frac{V_{,T}}{V}=0.  \label{KG}
\end{equation}%
The following potential was studied in \cite{tach0}:%
\begin{equation}
V(T)=\frac{\Lambda \sqrt{1-(1+k)y^{2}}}{1-y^{2}},  \label{pot}
\end{equation}%
with%
\begin{equation}
y=\cos \left[ \frac{3}{2}{\sqrt{\Lambda \,(1+k)}\,T}\right] ,  \label{pot1}
\end{equation}%
where $\Lambda >0$ and $-1<k<1$ are model parameters. The case $k>0$ is of
particular interest, because it reveals two unusual features: a
self-transformation of the tachyon into a pseudotachyon field and the
appearance of a Big Brake cosmological singularity. For $k>0$, the potential
(\ref{pot})--(\ref{pot1}), is well defined in the range%
\begin{equation}
-y_{\ast }<y<y_{\ast },\ \text{or}\ T_{4}>{T}>T_{3},
\end{equation}%
where 
\begin{equation}
y_{\ast }=(1+k)^{-1/2},
\end{equation}%
\begin{equation}
T_{3}=\frac{2}{3\sqrt{\Lambda (1+k)}}\arccos (1+k)^{-1/2},
\end{equation}%
\begin{equation}
T_{4}=\frac{2}{3\sqrt{\Lambda (1+k)}}\arccos \left[ \pi -(1+k)^{-1/2}\right]
.
\end{equation}%
Note that the dynamical system is invariant under the simultaneous change%
\begin{equation}
y\rightarrow -y\ ,\ \ \dot{T}\rightarrow -\dot{T}\ .  \label{inv}
\end{equation}%
Since $V\geq 0$ the pressure is negative allowing for an accelerated
expansion of the universe. When they reach the attractive critical point $%
(y=0,\dot{T}=0)$, the trajectories correspond to an exact de Sitter
expansion of the universe. The lines $\dot{T}=\pm 1$ (with the exception of
the corner points ($\pm y_{\ast }$, $\dot{T}=\pm 1$)) in the ($T$,$\dot{T}$)
space correspond to a standard Big Bang singularity (see Fig. 5 of \cite%
{tach2}, which reproduces Fig. 4 in \cite{tach0}).


However, some trajectories can reach the corner points where the geometry is
not singular. Hence, the trajectories can be continued across these corner
points, beyond which $|\dot{T}|^{2}$ becomes larger than $1$. The potential $%
V$ and the kinetic term in the Lagrangian density (\ref{L}) become imaginary
across the corner points; however their product remains real. Thus for $\dot{%
T}^{2}>1$ the correct Lagrangian density (describing a Born-Infeld type
pseudotachyon field) is 
\begin{equation}
L=W(T)\sqrt{g^{tt}\dot{T}^{2}-1},  \label{pseudo-Lagr}
\end{equation}%
where 
\begin{equation}
W(T)=\frac{\Lambda \sqrt{(1+k)y^{2}-1}}{1-y^{2}}.  \label{WT}
\end{equation}%
This Lagrangian is well defined in the ranges 
\begin{equation}
-1<y<-y_{\ast },\ \text{or }T_{\max }>{T}>T_{4}
\end{equation}%
and%
\begin{equation*}
y_{\ast }<y<1,\ \text{ or }T_{3}>{T}>0,
\end{equation*}%
with%
\begin{equation}
T_{\max }=\pi /3\sqrt{\Lambda (1+k)}.
\end{equation}%
The energy density and pressure are now 
\begin{equation}
\rho _{T}=\frac{W\left( T\right) }{\sqrt{\dot{T}^{2}-1}},  \label{rhoT}
\end{equation}%
\begin{equation}
p_{T}=W\left( T\right) \sqrt{\dot{T}^{2}-1}.  \label{pT}
\end{equation}%
Since the pressure is positive, the expansion of the universe is slowing
down. The field equation is%
\begin{equation}
\frac{\ddot{T}}{1-\dot{T}^{2}}+3H\dot{T}+\frac{W_{,T}}{W}=0.  \label{fieldEq}
\end{equation}

The universe runs into a soft singularity somewhere in the ranges $y_{\ast
}<y<1$ as ${\dot{T}\rightarrow -\infty }$ or $-1<y<-y_{\ast }$ as ${\dot{T}%
\rightarrow \infty }$ \cite{tach0}. From Eqs. (\ref{WT})-(\ref{pT}), the
potential $W$ is finite and $\rho _{T}\rightarrow 0$, $p_{T}\rightarrow
\infty $ at the soft singularity. Equivalently, this means that $H_{S}=0$,
while $\dot{H}_{S}=-\infty $. It was shown in \cite{tach2} that the
evolution of the universe can be continued across the singularity, where the
universe starts recollapsing and eventually ends in a Big Crunch singularity.

\subsection{Introducing a dust component\label{Intdust}}

The dust is a perfect fluid with vanishing pressure, whose energy density is 
\begin{equation}
\rho _{m}=\frac{\rho _{m,0}}{a^{3}},~  \label{dust}
\end{equation}%
where $\rho _{m,0}$ is a positive constant, characterizing the quantity of
matter in the universe today ($a_{0}=1$). Therefore, if a cosmological model
with dust evolves into a sudden singularity the energy density of dustlike
matter remains finite $\left( \rho _{m}\right) _{S}>0$. Then, the Hubble
parameter does not vanish at the singularity as in the case of the Big
Brake. This makes it more difficult and delicate to describe what happens
after reaching a soft singularity.

\subsubsection{Anti-Chaplygin gas}

A soft singularity arising in a two-fluid model containing dust and
anti-Chaplygin gas was investigated in \cite{Paradox}. The Hubble parameter
is positive at the singularity, requiring a further expansion of the
universe. Then a paradox arises: if the universe continues to expand, beyond
the singularity the expression under the sign of the square root in Eq. (\ref%
{solrhoach}) becomes negative and the energy density of the anti-Chaplygin
gas becomes ill defined.

A mathematically consistent way out of this situation is an abrupt
replacement of the cosmological expansion by a contraction at the price of
introducing distributional cosmological quantities \cite{Paradox}.

In the next section, we investigate an alternative possibility requiring the
smoothness in the evolution of the Hubble parameter but allowing for a
change in the equation of state (\ref{anti-Chap}).

\subsubsection{Born-Infeld type pseudotachyon field with trigonometric
potential}

In the model suggested in \cite{tach0} the Born-Infeld type pseudotachyon
field runs into a soft Big Brake singularity at some point during the
expansion of the universe. What happens however in the presence of a dust
component? Does the universe still run into a soft singularity?

In order to answer this question, we rewrite Eq. (\ref{fieldEq}) as 
\begin{equation}
\ddot{T}=(\dot{T}^{2}-1)\left( 3H\dot{T}+\frac{W_{,T}}{W}\right) .
\label{KG1}
\end{equation}%
In the left lower and in the right upper stripes (see Fig. 4 of \cite{tach0}%
), where the trajectories describe the expansion of the universe after the
transformation of the tachyon into the pseudotachyon field, the signs of $%
\ddot{T}$, of $\dot{T}$ and of the term $W_{,T}/W$ coincide. A detailed
analysis based on this fact was carried out in \cite{tach0} and led to the
conclusion that the universe encounters the singularity as $T\rightarrow
T_{S}$ ($T_{S}>0$ or $T_{S}>T_{max}$) , $|\dot{T}|\rightarrow \infty $. The
presence of dust cannot alter this because it increases the influence of the
term $3H\dot{T}$, and hence, accelerates the encounter with the singularity.
Indeed, consider two trajectories, crossing one of the corners (i.e.,
undergoing the tachyon-pseudotachyon transition) under the same angle in
phase space (cf. Fig. 4 in \cite{tach0}), one in the absence of dust, the
other in the presence of dust. For both trajectories the signs of $\ddot{T}$
and of $\dot{T}$ in (\ref{KG1}) coincide and the increase that the value of $%
H$ undergoes when dust is present makes the growing of $|\dot{T}|$ more
abrupt. On the other hand, the evolution of the tachyon field, approaching
the corner point is slowed down by the presence of dust, because, in this
case (inside the rectangle of the phase space) $(\dot{T}^{2}-1)$ is negative
and therefore $\ddot{T}$ and $\dot{T}$ have opposite signs. Summing up, we
may say that the presence of dust accelerates the evolution of the
pseudotachyon whereas it slows down the evolution of tachyon.

What is important is that the presence of dust changes in an essential way
the time dependence of the pseudotachyon field close to the singularity.
Indeed, as it was shown in \cite{tach2}, in the absence of dust one has 
\begin{equation}
T=T_{BB}\pm \left( \frac{4}{3W(T_{BB})}\right) ^{1/3}(t_{BB}-t)^{1/3},
\label{old-asymp}
\end{equation}%
(see Eq. (29) in \cite{tach2}). The upper (lower) sign corresponds to the
left lower (right upper) strip in Fig. 4 of \cite{tach0}, where $%
\lim_{t\rightarrow t_{BB}}\dot{T}=-\infty $ ( $\lim_{t\rightarrow t_{BB}}%
\dot{T}=\infty $). In the presence of dust one has, instead, 
\begin{equation}
T=T_{S}\pm \sqrt{\frac{2}{3H_{S}}}\sqrt{t_{S}-t},  \label{new-asymp}
\end{equation}%
where $H_{S}$ is the nonvanishing value of the Hubble parameter given by 
\begin{equation}
H_{S}=\sqrt{\frac{\rho _{m,0}}{a_{S}^{3}}}.  \label{H-S}
\end{equation}%
Here we have taken advantage of the fact that in Eq. (\ref{KG1}) the terms $%
1 $ and $W_{,T}/W$ can be neglected with respect to $\dot{T}^{2}$ and $3H%
\dot{T}$, respectively. It is easy to see that a smooth continuation of
expression (\ref{new-asymp}) is impossible in contrast to the situation
without dust (\ref{old-asymp}).

Thus, the presence of dust is responsible for the appearance of similar
paradoxes in both the anti-Chaplygin gas and tachyon models.

\section{Crossing the soft singularity and transformations of matter \label%
{smoothCross}}

As mentioned earlier (see Introduction of the present paper and the
concluding remarks in \cite{Paradox}) the mathematically self-consistent
scenario, based on the treatment of physical quantities as generalized
functions and on the abrupt change of the expansion into a contraction, may
look counterintuitive from the physical point of view. Indeed, such a
behavior displays features which are analogous to the phenomenon of the
absolutely elastic bounce of a hard ball from a rigid wall, as studied in
classical mechanics. In the latter case, it is the velocity and the momentum
of the ball which change their direction abruptly. Hence, an infinite force
acts from the wall onto the ball during an infinitely small interval of time.

In reality, the absolutely elastic bounce is an idealization of a process
taking place in a finite, though small, time-span, during which inelastic
deformations of the ball and of the wall occur. This implies a more complex
and realistic description of the dynamical process of interaction between
the ball and the wall. Hence, we are naturally led to assume that something
similar should occur also in the models of an anti-Chaplygin gas or a
tachyon whenever dust is present. We expect that the smoothing of the
process of the transition from an expanding to a contracting phase should
include some (temporary) geometrically implied change of the equation of
state of matter or of the form of the Lagrangian. We know that such changes
have been considered in cosmology. For example, a tachyon--pseudo-tachhyon
transformation, driven by the continuity of the cosmological evolution, took
place in the tachyon model \cite{tach0} (see also subsection II.B.2 of the
present paper). In a cosmological model with the phantom field with a cusped
potential \cite{cusped}, transformations between phantom and standard scalar
field were considered. Thus, it is quite natural to assume that the process
of crossing of the soft singularity should imply similar transformations.

However, the situation is now more complicated. It is not enough to require
the continuity of the evolution of the cosmological radius and of the Hubble
parameter. It is also necessary to make some hypotheses about changing the
equation of state of matter or the form of the Lagrangian.

We solve the problem as follows. Considering first the anti-Chaplygin gas
with dust, we require a minimal change in the form of the dependence of the
energy density and of the pressure on the cosmological radius, upon crossing
the soft singularity. This will require replacement of the anti-Chaplygin
gas with a Chaplygin gas with negative energy density\footnote{%
A Chaplygin gas with negative energy density has been considered earlier 
\cite{Khalat} in a different context.}. Next, we consider the cosmological
model based on a pseudotachyon field with constant potential and in the
presence of dust. It is known that the energy-momentum tensor for such a
pseudotachyon field coincides with that of the anti-Chaplygin gas (relating
the Chaplygin gas to the tachyon field with constant potential was
considered in \cite{FKS}). We derive how the pseudotachyon Lagrangian
transforms using its kinship with the anti-Chaplygin gas. In this way, we
arrive at a new type of Lagrangian, belonging to the \textquotedblleft
Born-Infeld family\textquotedblright . Finally, we extend this
transformation to the case of the trigonometric potential.

\subsection{Anti-Chaplygin gas}

It follows from Eqs. (\ref{anti-Chap}) and (\ref{solrhoach}) that the
pressure of the anti-Chaplygin gas 
\begin{equation}
p = \frac{A}{\sqrt{\frac{B}{a^6}-A}}  \label{pressure-anti}
\end{equation}
tends to $+\infty$ when the universe approaches the soft singularity, e.g.
when the cosmological radius $a \rightarrow a_S$ (see Eq. (\ref{radius-sing}%
)). Requiring the expansion to continue into the region $a > a_S$, while
changing minimally the equation of state, we assume 
\begin{equation}
p = \frac{A}{\sqrt{|\frac{B}{a^6}-A|}},  \label{pressure-new}
\end{equation}
or, in other words, 
\begin{equation}
p = \frac{A}{\sqrt{A-\frac{B}{a^6}}},\ \mathrm{for}\ a > a_S.
\label{pressure-new1}
\end{equation}
Thus, the pressure passes through $+\infty$ conserving its sign, thus
providing in such a way the continuity of the cosmological evolution. 
It is crucial that $p$ does not change sign in order to keep a decelerated
expansion. 
The energy density $\rho$ evolves continuously, and so does its derivative
with respect to volume. 
Combining (\ref{pressure-new1}) with the energy conservation law (\ref{cont}%
) we obtain 
\begin{equation}
\rho = -\sqrt{A-\frac{B}{a^6}},\ \mathrm{for}\ a > a_S,  \label{energy-new}
\end{equation}
so that for $a > a_S$ the energy density and the pressure satisfy the
Chaplygin gas equation of state 
\begin{equation}
p = -\frac{A}{\rho}.  \label{Chaplygin}
\end{equation}
Therefore, at the singularity crossing, the anti-Chaplygin gas transforms
into a Chaplygin gas with negative energy density. After crossing of the
singularity the Friedmann equation is 
\begin{equation}
H^{2}=\frac{\rho _{m,0}}{a^{3}}-\sqrt{A}\sqrt{1-\left( \frac{a_{S}}{a}%
\right) ^{6}},  \label{Friedmann2}
\end{equation}%
and it follows from Eq. (\ref{Friedmann2}) that, after achieving the point
of maximal expansion $a = a_{\mathrm{max}}$, where 
\begin{equation}
a_{\mathrm{max }}=\left( \frac{\rho _{m,0}^{2}}{A}+a_{S}^{6}\right) ^{1/6},
\label{astar}
\end{equation}%
the universe begins contracting. During this phase, as it achieves again $%
a=a_S$, it stumbles once more upon a soft singularity, whereupon the
Chaplygin gas transforms itself back into anti-Chaplygin with positive
energy density and the contraction continues until hitting the Big Crunch
singularity.

Whereas in \cite{Paradox} we envisaged an abrupt change from expansion to
contraction through the singularity, with a jump in the Hubble parameter, we
show here that a continuous transition to the collapsing phase is possible
if the equation of state of the anti-Chaplygin gas has some kind of ``phase
transition'' at the singularity.

\subsection{Pseudotachyon field with a constant potential}

For a pseudotachyon field with constant potential $W(T)=W_{0}$, the energy
density (\ref{rhoT}) and the pressure (\ref{pT}) satisfy the anti-Chaplygin
gas equation of state (\ref{anti-Chap}) with 
\begin{equation}
A=W_{0}^{2}.  \label{WA}
\end{equation}%
Solving the equation of motion for the pseudotachyon field (\ref{fieldEq})
with $W(T)=W_{0}$, one finds 
\begin{equation}
\dot{T}^{2}=\frac{1}{1-\left( \frac{a}{a_{S}}\right) ^{6}}  \label{Tdot1}
\end{equation}%
and we see that a soft singularity arises at $a=a_{S}$ with $\dot{T}%
^{2}\rightarrow +\infty $.

The new Lagrangian, which gives the correct energy density and pressure
satisfying a Chaplygin gas equation with negative energy density is 
\begin{equation}
L=W_{0}\sqrt{g^{tt}\dot{T}^{2}+1},~~~~~~~~~~~~~~~~~a>a_{S}  \label{BI-new}
\end{equation}%
giving 
\begin{equation}
p=W_{0}\sqrt{\dot{T}^{2}+1}  \label{pressure-BI}
\end{equation}%
and 
\begin{equation}
\rho =-\frac{W_{0}}{\sqrt{\dot{T}^{2}+1}}.  \label{energy-BI}
\end{equation}%
Lagrangian (\ref{BI-new}) characterizes a new type of Born-Infeld field,
which we may call \textquotedblleft quasitachyon\textquotedblright .

For an arbitrary potential the Lagrangian reads 
\begin{equation}
L=W(T)\sqrt{g^{tt}\dot{T}^{2}+1}~~~~~~~~~~~~~~~~~a>a_{S}  \label{Lagr-new}
\end{equation}%
with equation of motion 
\begin{equation}
\frac{\ddot{T}}{\dot{T}^{2}+1}+3H\dot{T}-\frac{W_{,T}}{W}=0  \label{KG-BI}
\end{equation}%
and energy density and pressure are, respectively, 
\begin{equation}
\rho =-\frac{W(T)}{\sqrt{\dot{T}^{2}+1}}~  \label{energy-BI1}
\end{equation}%
and 
\begin{equation}
p=W(T)\sqrt{\dot{T}^{2}+1}.
\end{equation}%
If $W(T)=W_{0}$, the solution of equation (\ref{KG-BI}) is 
\begin{equation}
\dot{T}^{2}=\frac{1}{\left( \frac{a}{a_{S}}\right) ^{6}-1},  \label{Tdot2}
\end{equation}%
and the energy density evolves as%
\begin{equation}
\rho _{T}=-W_{0}\sqrt{1-\left( \frac{a_{S}}{a}\right) ^{6}}.  \label{rho-BI}
\end{equation}%
The evolution of the universe coincides with that of a universe with
anti-Chaplygin gas and dust.

Thus, the transformation from anti-Chaplygin to Chaplygin gas with negative
energy density corresponds to a transition from a pseudotachyon field with
Lagrangian (\ref{pseudo-Lagr}) with constant potential $W(T)=W_{0}$ to a new
Born-Infeld type quasitachyon field, with Lagrangian (\ref{BI-new}).

\subsection{The tachyon model with trigonometric potential and dust}

In the vicinity of the soft singularity, it is the \textquotedblleft
friction\textquotedblright\ term $3H\dot{T}$ in the equation of motion (\ref%
{fieldEq}), which dominates over the potential term $W_{,T}/W$. Hence, the
dependence of $W(T)$ on its argument is not essential and a pseudotachyon
field approaching this singularity behaves like one with a constant
potential. Thus, it is reasonable to assume that upon crossing the soft
singularity the pseudotachyon transforms itself into a quasitachyon with
Lagrangian (\ref{Lagr-new}) for any potential $W(T)$.

We now study the dynamics of the model with trigonometric potential (\ref%
{pot})-(\ref{pot1}) in the presence of dust.

The behavior of the quasitachyon field close to the soft singularity can be
derived from Eq. (\ref{KG-BI}) in the same way as the corresponding behavior
of the pseudotachyon field derives from Eq. (\ref{KG1}). In analogy with Eq.
(\ref{new-asymp}), we obtain the quasitachyon behavior 
\begin{equation}
T=T_{S}\mp \sqrt{\frac{2}{3H_{S}}}\sqrt{t-t_{S}}
\end{equation}%
and the two formulas match with each other through the singularity.

In order to analyze the dynamics of the field in the presence of dust, it is
convenient to concentrate ourselves on the processes as they occur, say, in
the left lower strip of the phase diagram of the model, to facilitate
comparison with earlier studies of the tachyon model dynamics without dust
in \cite{tach0,tach2}. The relative signs in the equations of motion of the
term with the second derivative $\ddot{T}$ and of the friction term $3H\dot{T%
}$ are opposite for pseudotachyons and quasitachyons. This means that after
crossing the soft singularity the time derivative $\dot{T}$ grows while its
absolute value decreases. At the same time the value of $T$ is decreasing
while the potential $W(T)$, given by (\ref{WT}) is growing.

Hence the absolute value of the negative contribution to the energy density
of the universe induced by the quasitachyon grows while the energy density
of the dust decreases due to the expansion of the universe. Thus, at some
moment the total energy density vanishes and the universe reaches the point
of maximal expansion, after which the expansion is replaced by a contraction
and the Hubble variable changes sign. The change of sign of the friction
term $3H\dot{T}$ implies the value of $\dot{T}$ to decrease and at some
finite moment of time the universe hits again the soft singularity when $%
\dot{T}\rightarrow -\infty $. Upon crossing this singularity the
quasitachyon transforms back to pseudotachyon and the relative signs of the
terms with the second and first time derivatives in the equation of motion
change once again. After this, the time derivative of the pseudotachyon
field begins to grow and the universe continues its contraction until it
hits the Big Crunch singularity.

It was shown in \cite{tach2} that, for the case of the tachyon model with
trigonometric potential and without dust, the encounter of the universe with
the Big Crunch singularity occurs at $T=0$ and $\dot{T}=-\sqrt{\frac{1+k}{k}}
$. One can show that the presence of dust does not change these values.
Indeed, consider the behavior of the pseudotachyon field when $T\rightarrow
0,\ \dot{T}\rightarrow -\sqrt{\frac{1+k}{k}}$. It follows from the
expressions (\ref{rhoT}) and (\ref{pT}) that the ratio between pressure and
energy density behaves as 
\begin{equation}
\frac{p}{\rho }=\dot{T}^{2}-1\rightarrow \frac{1}{k},  \label{barotropic}
\end{equation}%
i.e. in the vicinity of the Big Crunch singularity the pseudotachyon field
behaves as a barotropic fluid with the equation of state parameter $\frac{1}{%
k}>1$. This means that the energy density of the pseudotachyon field grows
as 
\begin{equation}
\rho \sim \frac{1}{a^{3(1+\frac{1}{k})}}  \label{rad-sing}
\end{equation}%
with $a\rightarrow 0$, namely much more rapidly than the dust energy
density. Thus, one can neglect the contribution of dust in the regime of
approach to the Big Crunch singularity and the description of the evolution
of the universe to this point coincides with that of the pure tachyon model 
\cite{tach2}.

\subsection{Additional remarks concerning geometrically induced
transformations of matter properties}

Before addressing the numerical study of the cosmological evolutions in the
tachyon model with trigonometrical potential, we would like to dwell on some
basic features of the matter transformations introduced in this section.

Concerning the transformation from the anti-Chaplygin gas with the equation
of state (5) to the Chaplygin gas with the equation of state (36), we would
like to emphasize that this is not an extension of the definition of the
anti-Chaplygin gas into the region, where it was not defined before, but
instead that it is a transition from one perfect fluid into another one.
This transition is the result of a complicated interplay between the
evolution of the spacetime described by the Friedmann equations and the
evolution of perfect fluids, described by the continuity equations. Indeed,
in the description of this transition we use not only the equations of state
of fluids, but also the explicit dependences of their energy densities and
pressure on the cosmological radius. Thus, in describing the passage of the
universe filled with the anti-Chaplygin gas and with dust through the soft
singularity, we put forward two requirements: first, the cosmological
evolution should be as smooth as possible; second, the change of the
character of the dependence of the energy density and of the pressure of the
fluids should be minimal. These two requirements imply the substitution of
the formula (32) giving the pressure of the anti-Chaplygin gas in the
vicinity of the singularity with the formula (33), yielding Eq. (34) for $a
> a_{S}$. Such a substitution provides the conservation of the sign of the
pressure and the smoothness of the cosmological evolution. After that the
continuity equation (3) gives the expression (35) for the energy density of
the fluid and we easily see that the anti-Chaplygin gas has been transformed
into a Chaplygin gas with a negative energy density.

The situation with transformations of the tachyon field is more complicated.
First of all, let us note that there are two different kinds of
transformations, the transformation from tachyon to pseudotachyon and the
transformation from pseudotachyon to quasitachyon. The first kind of
transformation was introduced in the paper [2] and it is the transformation
of the field with the Lagrangian (8) and the potential (13) into the field
with the Lagrangian (20) and the potential (21). This transformation is not
connected with the crossing of the singularity. When the pressure of the
tachyon field vanishes, the potential and kinetic terms in the Lagrangian
(8) become ill defined. However, the equations of motion of this field can
be continued to the part of the phase space of the corresponding dynamical
system, where the pressure is positive. The new Lagrangian (20), (21), well
defined in this region, gives the equation of motion which coincides with
the old equation of motion given by the Lagrangian (8), (13). Formally, we
can describe this transition by introducing the absolute values into the
expressions under the square root sign in both the kinetic and potential
terms of the Lagrangian (8). However, we would like to stress that the main
role in the transformation from the tachyon to the pseudotachyon is played
by the equations of motion.

The justification of the transition from the pseudotachyon field to the
quasitachyon field with the Lagrangian (44) is more subtle. This
transformation is induced by the crossing of the soft singularity in the
presence of dust and there is no way to use the continuity of the form of
the Lagrangian or the conservation of the form of the equations of motion.
We use instead the fact that the equation of state of the pseudotachyon
field with constant potential coincides exactly with that of the
anti-Chaplygin gas. Thus, to provide a passage which is as smooth as
possible of the universe filled with the pseudotachyon field with constant
potential through the soft singularity we should find such a Lagrangian of a
Born-Infeld type field which is equivalent to the Chaplygin gas with a
negative energy density. Following this path we come to the quasitachyon
field with Lagrangian (41). The last step consists in the generalization of
the Lagrangian (41) for the case of an arbitrary potential (44). Such a
generalization is justified by the fact, that in the vicinity of the soft
singularity, the change of the potential term of the pseudotachyon field is
much slower in comparison with the kinetic term.

While the transition from the pseudotachyon to the quasitachyon is more
radical and intricated than the other matter transformations considered here
and in the preceding papers, it still looks quite logical and probably the
only one which is possible.

The construction developed in the paper might be interpreted as gluing two
charts of a Friedmannian universe across the (spatially homogeneous)
hypersurfaces of singularity. In the case of the fluid, its energy density
is positive in one chart and negative in the other chart, with separate
forms of equations of state in each chart. As a homogeneous universe is an
idealization, let us conclude this subsection with a remark concerning the
possible generalization to inhomogeneous cosmologies. Here the gluing can
still be enforced along the hypersurface with zero energy density of the
exotic fluid. For the scalar field the gluing hypersurface could be also
defined as having zero energy density, however its definition would be more
cumbersome, due to the different Lagrangians of the field in the two charts.

\section{Future evolution of the tachyon field with trigonometric potential
and dust: numerical results}

The tachyon model with trigonometric potential was tested in \cite{tach1} by
comparing it with SNIa data. In that paper we found the range of values of
the model parameter $k$ and tachyon field initial conditions fitting well
the SNIa data. Then we studied future evolutions starting from acceptable
initial conditions. While a subset of the corresponding trajectories leads
to a de Sitter expansion, a complementary subset of trajectories leads to a
Big Brake singularity. The evolution after the Big Brake singularity
crossing was described in \cite{tach2}.

\begin{widetext}

\begin{figure}[ht]
\includegraphics[height=5cm, angle=0]{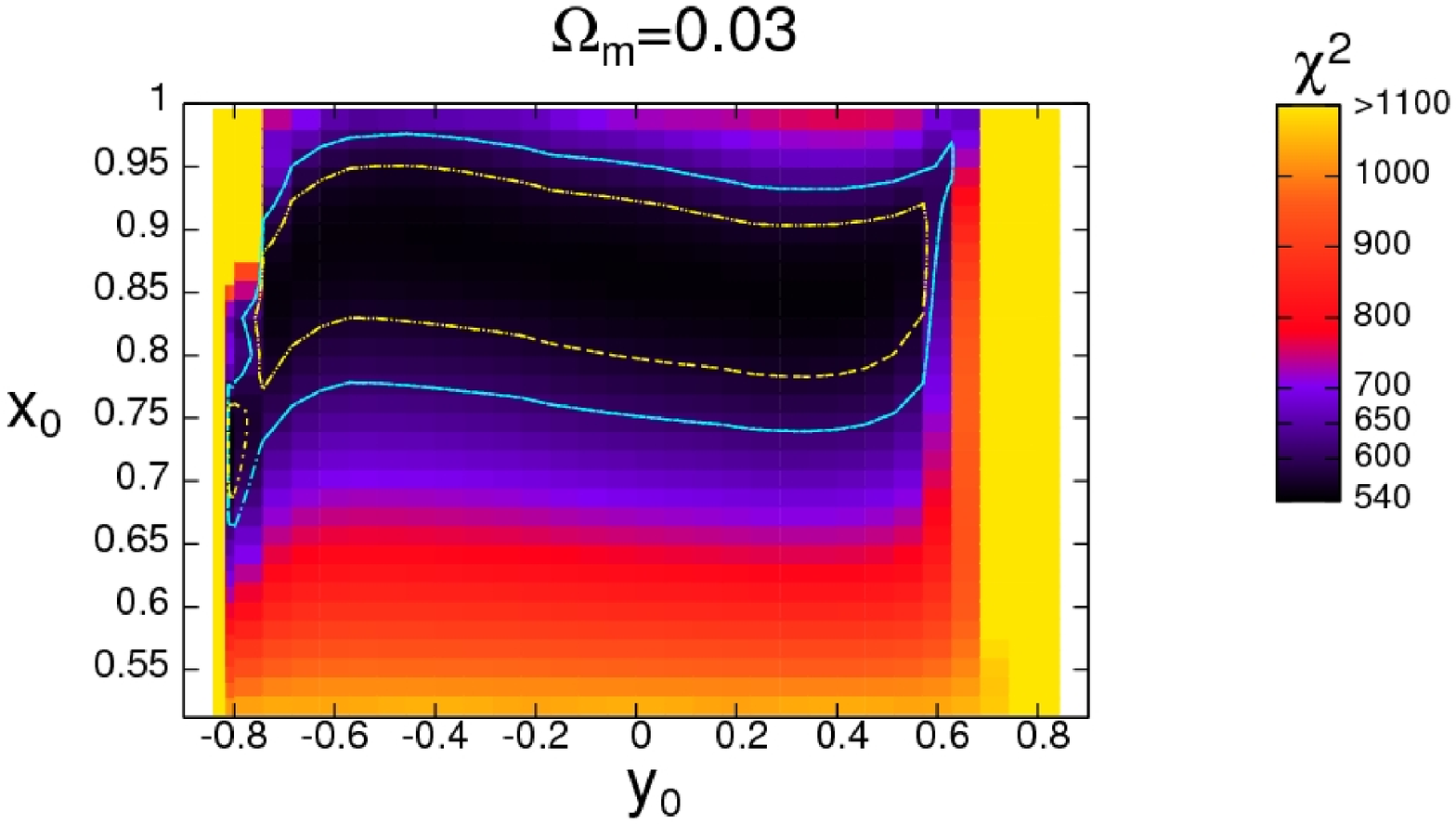}
\includegraphics[height=5cm, angle=0]{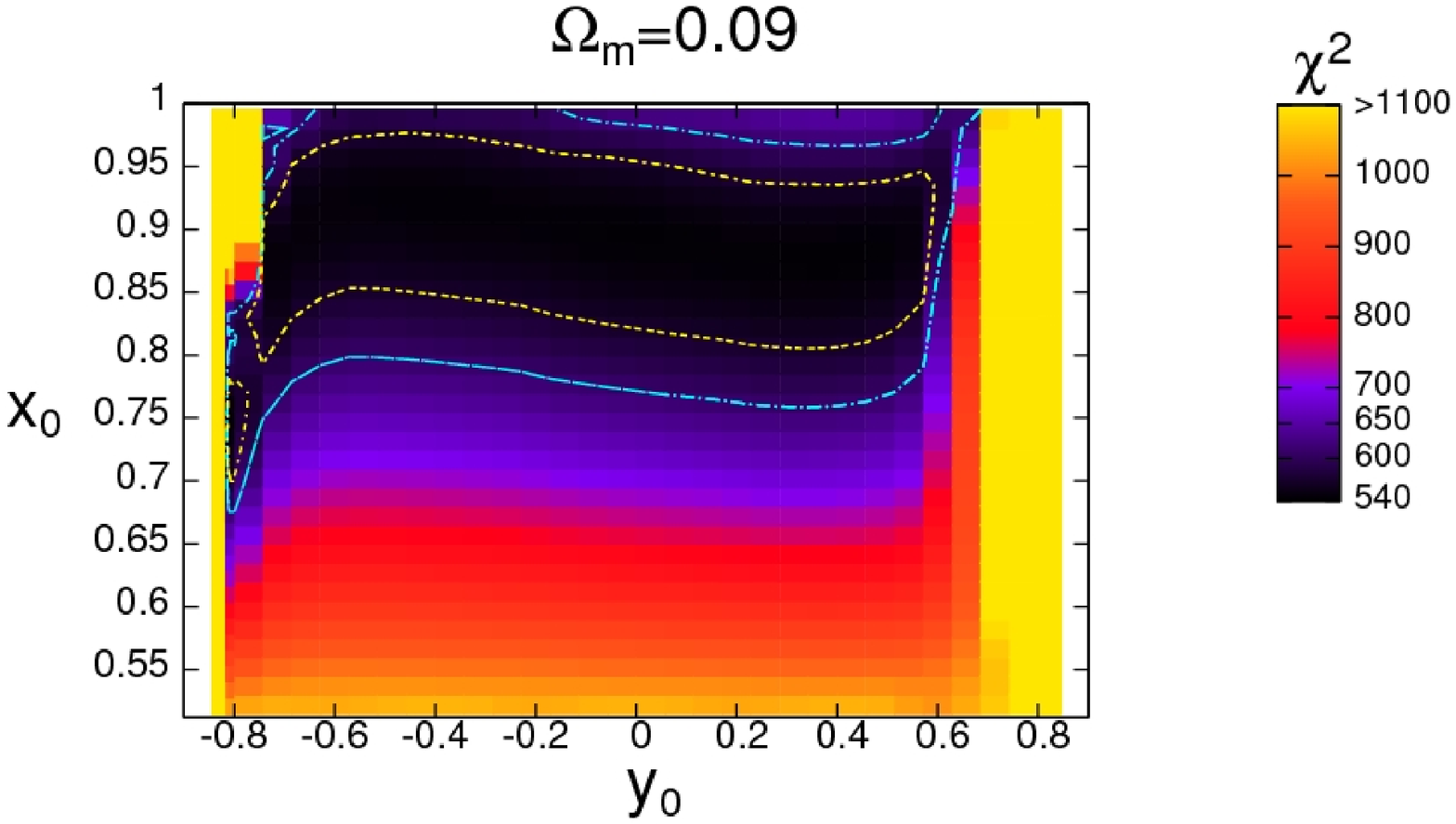}
\newline
\vskip 0.3cm
\includegraphics[height=5cm, angle=0]{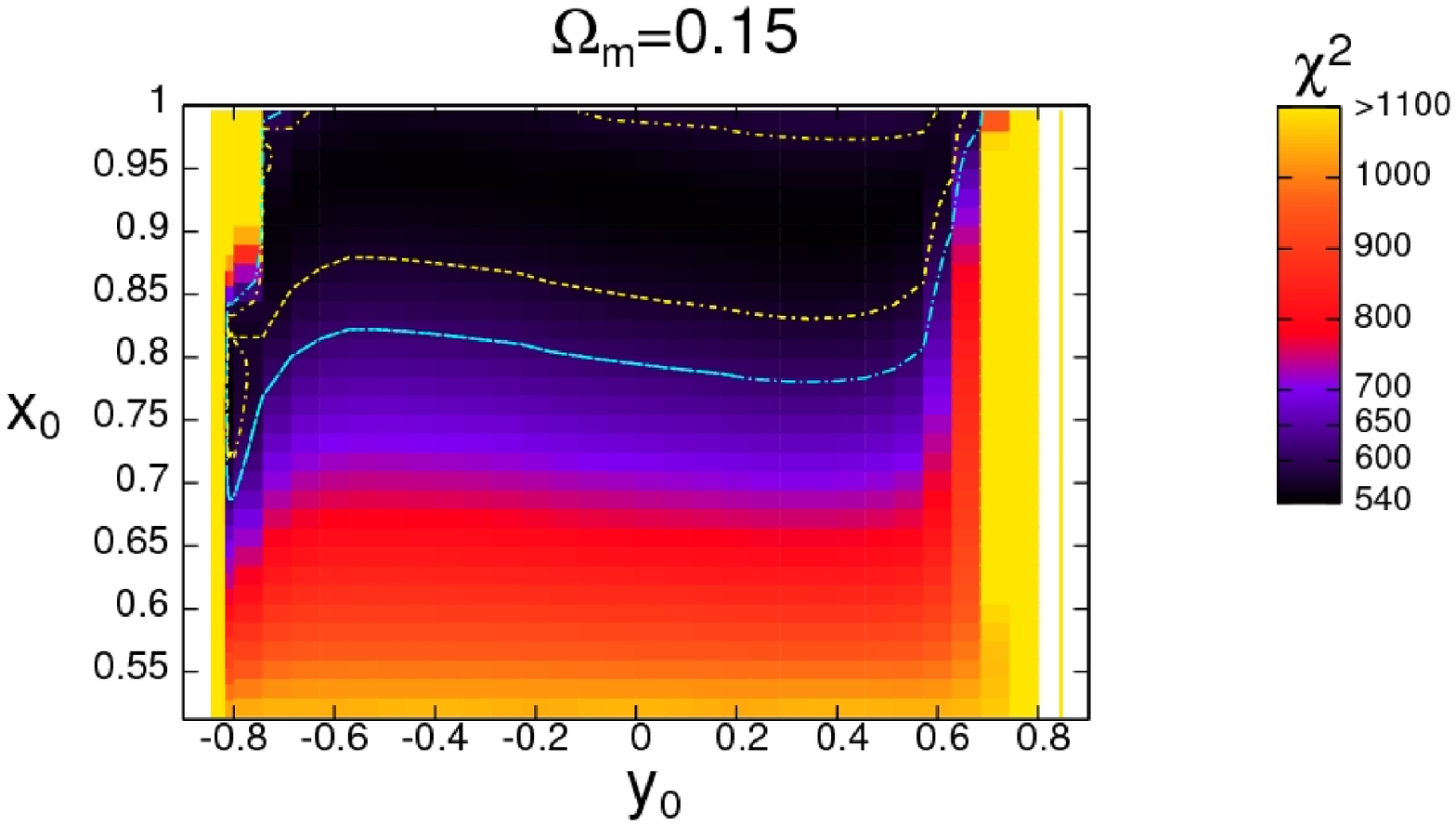}
\includegraphics[height=5cm, angle=0]{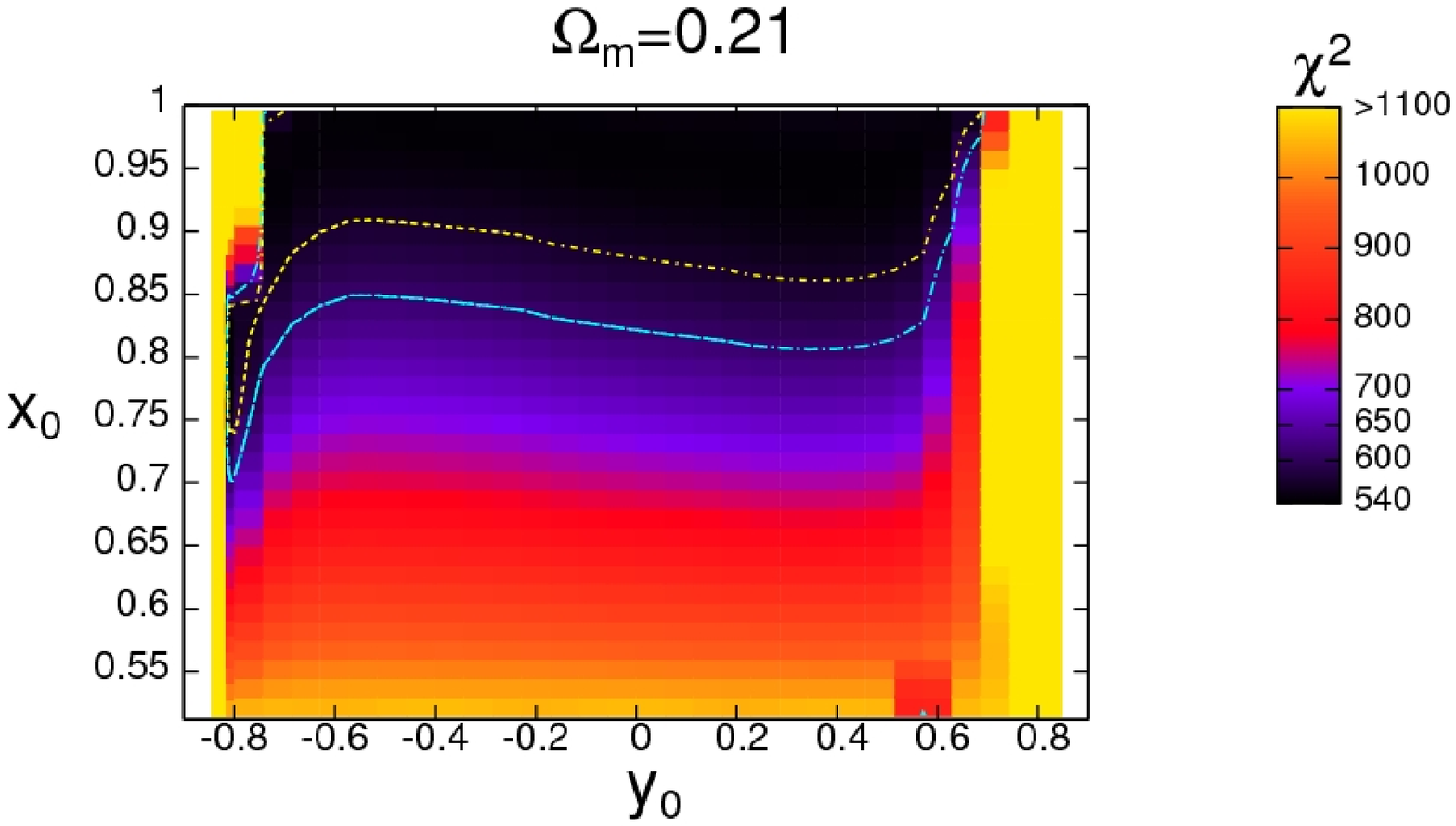}
\newline
\vskip 0.3cm
\includegraphics[height=5cm, angle=0]{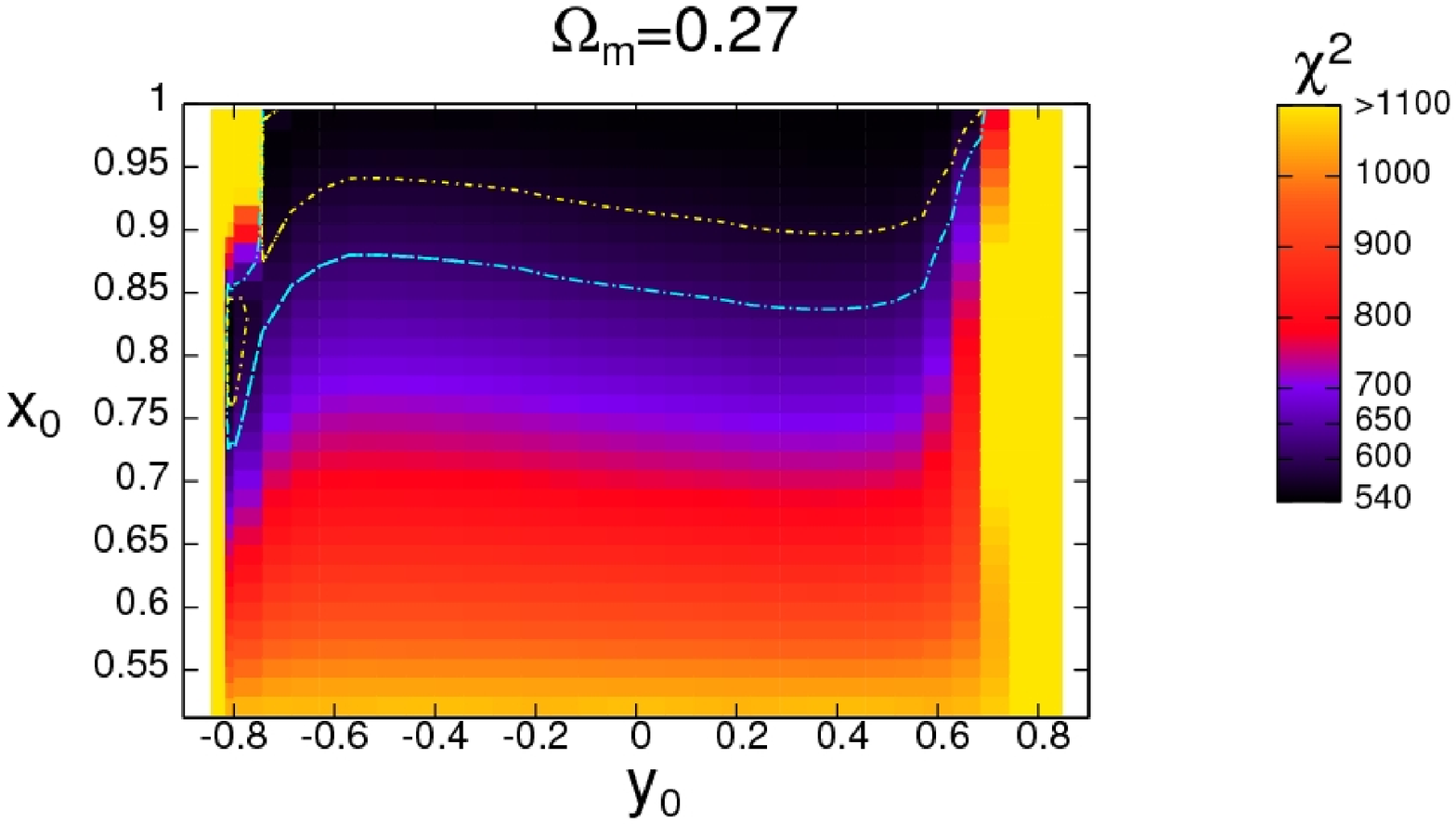}
\includegraphics[height=5cm, angle=0]{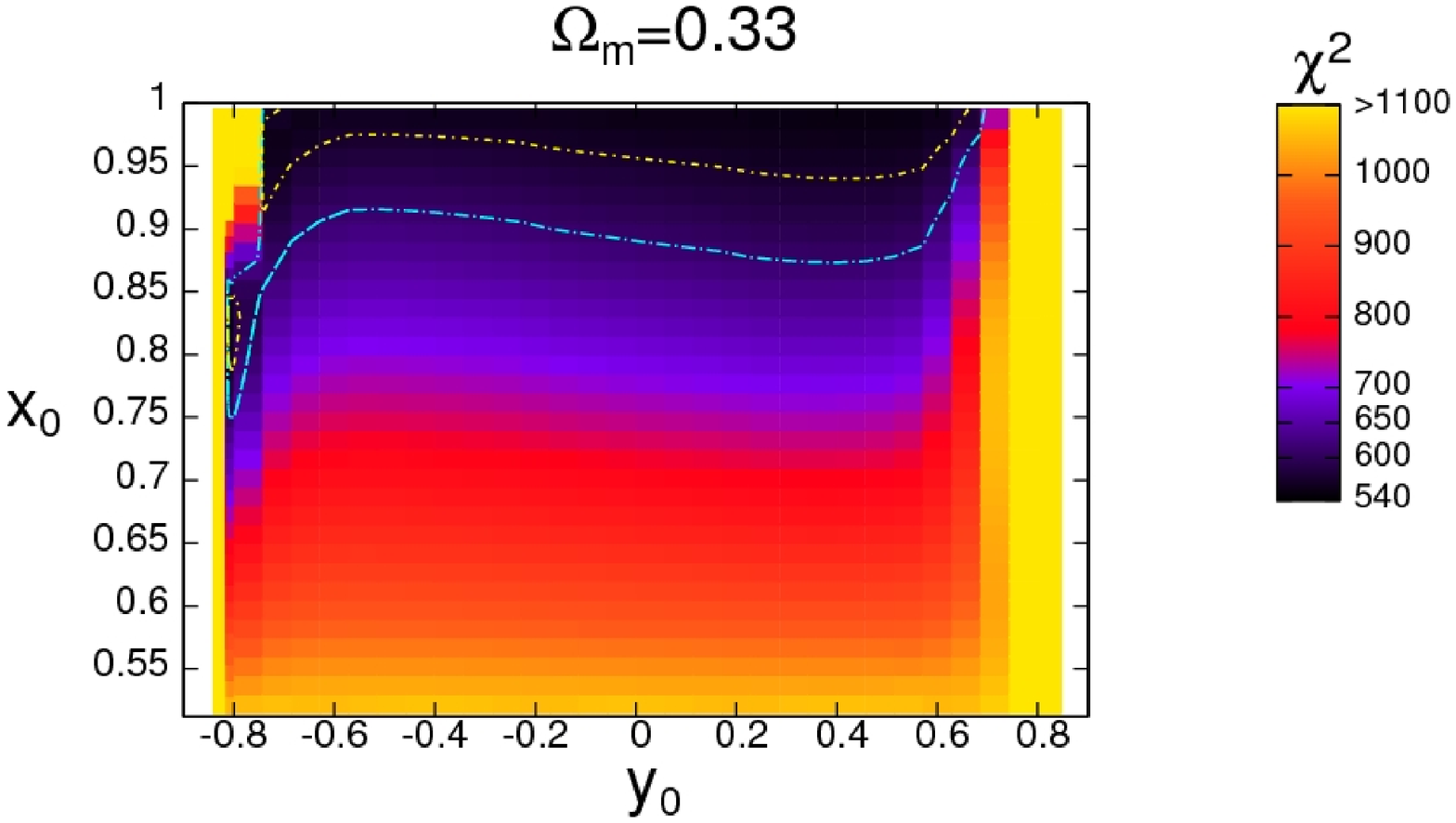}
\newline
\vskip 0.3cm
\caption{(Color online)
The fit of the luminosity distance vs redshift for the parameters $k=0.4$ and
$\Omega_m=0.03$ (upper left),
$0.09$ (upper right), $0.15$ (middle left), $0.21$ (middle right), $0.27$ (lower left), $0.33$ (lower right),
in the parameter plane ($y_{0}$, $x_{0}$) in the range $\left\vert y_{0}\right\vert \leq 0.845$ where the potential $V$ is well defined.
The contours refer to the $68.3\%$ (1$\sigma$) and $95.4\%$ (2$\sigma$) confidence
levels.
The color code for $\chi^2$ is indicated on the vertical stripes.
The clear tendency with increasing the dust component is that the parameter
$x_0$ approaches its maximally allowed value (representing $\dot{T}_0=0$). Higher values of $\Omega_m$ render the fit
with the supernovae outside the 1$\sigma$ region.
}
\label{Fig1}
\end{figure}

\end{widetext}

In subsection IVA we investigate the compatibility with SNIa data of this
dark energy model in the presence of dustlike matter. We use the Union2 SNIa
data set \cite{Union2}. We show that the model fits the SNIa data well also
in the presence of dustlike matter.

In subsection IVB we investigate the future evolution numerically for those
trajectories which run into the soft singularity at time $t_{S_{1}}$. We
give specifically the time intervals measured from today for the following
events: $t_{dec}$ when the cosmic expansion becomes decelerated; $t_{\ast }$
corresponding to the tachyon-pseudotachyon transformation (crossing the
corner of the rectangle in the phase portrait of the model); $t_{S_{1}}$
when the first soft singularity is reached; $t_{turn}$ corresponding to the
turning point when the universe starts contracting; $t_{S_{2}}$ when the
second soft singularity is reached; and finally the time $t_{BC}$ of the Big
Crunch.

\subsection{Test with supernovae data\label{SNIatest}}

The tachyon field violates the strong energy condition when $\dot{T}^{2}<1$,
as required by a dark energy candidate. 
For a reasonable fit with supernova data we assume $\dot{T}^{2}<1$. In the
regime where $a$ varies monotonically with time it may be convenient to
replace the cosmological time with a monotonic function of the scale factor
as a new independent variable. 
We choose the redshift $1+z=a_{0}/a$ as a new independent variable\footnote{%
We note that since the Friedmann equation is a first integral it can be used
as a check of the accuracy of the numerical integration.} (here and
henceforth the subscript $0$ refers to the value of the respective
quantities at the present epoch).

The model depends on the parameters $k$ and $\Lambda $. For given values of
these parameters the possible solutions depend on the quantity of dust and
on the initial conditions $y_{0},\dot{T}_{0}$ for the tachyon field.
However, the Friedmann equation implies the following constraint 
\begin{equation}
\dot{T}_{0}=\pm \sqrt{1-\frac{\ \Omega _{\Lambda ,0}^{2}\left[ 1-\left(
1+k\right) y_{0}^{2}\right] }{\left( 1-\Omega _{m,0}\right) ^{2}\left(
1-y_{0}^{2}\right) ^{2}}},
\end{equation}%
where (remembering our convention $8\pi G/3=1$)%
\begin{equation}
\,\Omega _{\Lambda ,0}=\frac{\Lambda }{H_{0}^{2}},~\Omega _{m,0}=\frac{\rho
_{m,0}}{H_{0}^{2}},  \label{new-var}
\end{equation}%
showing that $\Lambda $ is determined by the values of the other parameters.
In what follows we fix $k=0.4$ and vary $\Omega _{m,0}$ through the values $%
\{0.03$, $0.09$, $0.15$, $0.21$, $0.27$, $0.33\}$. As in paper \cite{tach1}
we avoid the double coverage of the parameter space (the model has a
symmetry given by Eq. (\ref{inv})) by replacing $\dot{T}_{0}$ \cite{tach1}
with the new variable\footnote{%
The parameter $x_{0}$ is denoted by $w_{0}$ in \cite{tach1}.}: 
\begin{equation}
x_{0}=\frac{1}{1+\dot{T}_{0}^{2}}.  \label{w0}
\end{equation}%
The initial conditions $x_{0}$ and $y_{0}$ vary inside the rectangle $\frac{1%
}{2}\leq x_{0}\leq 1,|y_{0}|\leq 0.845$. Finally, we introduce the
luminosity distance $d_{L}$ whose evolution is given by 
\begin{equation}
\frac{d}{dz}\left( \frac{d_{L}}{1+z}\right) =\frac{1}{H}.  \label{dLz}
\end{equation}

Fitting to the supernovae data involves a $\chi ^{2}$-test, as described in
Refs. \cite{DicusRepko}, \cite{tach1}. In Fig \ref{Fig1} we show the $\chi
^{2}$ values in the parameter plane of the initial conditions $\left(
y_{0},x_{0}\right) $. The contours correspond to the 1 $\sigma $,
respectively 2$\sigma $, confidence levels with $\chi ^{2}=570.34$ and $\chi
^{2}=612.33$, respectively.

\subsection{Future evolution\label{future}}

As in the preceding papers \cite{tach1,tach2} we study numerically the
future evolution of the universe starting with initial conditions compatible
with SNIa data. However, our task now is technically more complicated due to
the presence of dust. As a matter of fact, we shall have to consider five
different regimes, where different systems of dynamical equations are used
and we should provide four accurate matching between these evolutions.
First, the universe starts its evolution at some point in the rectangle on
the phase space of Fig. 4 of \cite{tach0}. Here the field $T$ satisfies the
equation of motion (\ref{KG}) and the right-hand side of the first Friedmann
equation includes the contribution of dust (\ref{dust}) and of the tachyon
field (\ref{rho}). After the crossing of the corner (at $t_{\ast }$), the
tachyon field transforms into a pseudotachyon field with equation of motion (%
\ref{fieldEq}) and energy density (\ref{rhoT}). This is the second regime.
The third regime enters into action after the first crossing of the soft
singularity (at $t_{S_{1}}$), when the pseudotachyon transforms itself into
a quasitachyon with equation of motion (\ref{KG-BI}) and energy density (\ref%
{energy-BI1}). After the passing of the point of maximal expansion of the
universe (at $t_{turn}$) we enter into the fourth regime when the universe
starts contracting. After the second soft singularity crossing (at $%
t_{S_{2}} $) we have the fifth regime, where the quasitachyon converts
itself again into a pseudotachyon. Finally, the universe ends in a Big
Crunch (at $t_{BC}$). The corresponding times are shown in Table \ref{Table1}
for $\Omega _{m,0}=0.03$ and in Table \ref{Table2} for $\Omega _{m,0}=0.27$ .

These times have been computed assuming for $H_{0}$ the value $70$ km s$%
^{-1} $ Mps $^{-1}$. It is known that there is a certain discrepancy between
the value of the Hubble parameter arising indirectly from the cosmic
microwave background and baryon acoustic oscillations \cite{Planck}, and the
one more directly obtained from local measurements of the relation between
redshifts and distances to sources \cite{Riess} (for a recent analysis of
this problem see \cite{variance}). The former gives $H_{0}^{CMB}=67.89\pm
0.77$ km s$^{-1} $ Mps$^{-1}$, while the latter gives $H_{0}^{local}=73.8\pm
2.4$ km s$^{-1}$ Mps$^{-1}$. Nevertheless, the precise value of $H_{0}$ is
not so important for our study, hence, we have taken an intermediate value.

\begin{widetext}

\begin{figure}[ht]
\vskip 0.4cm
\includegraphics[height=8cm, angle=270]{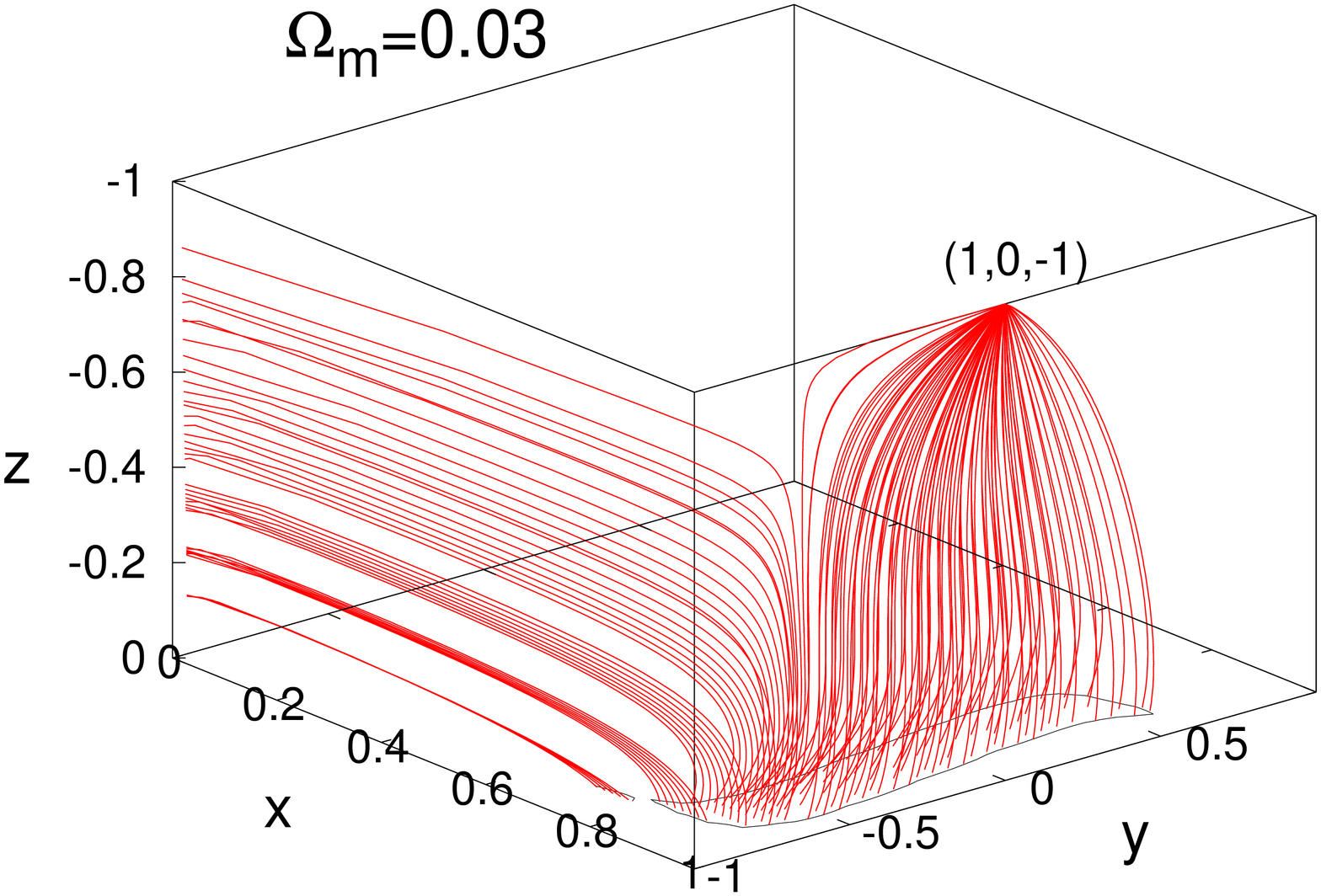}
\hskip 0.4cm
\includegraphics[height=8cm, angle=270]{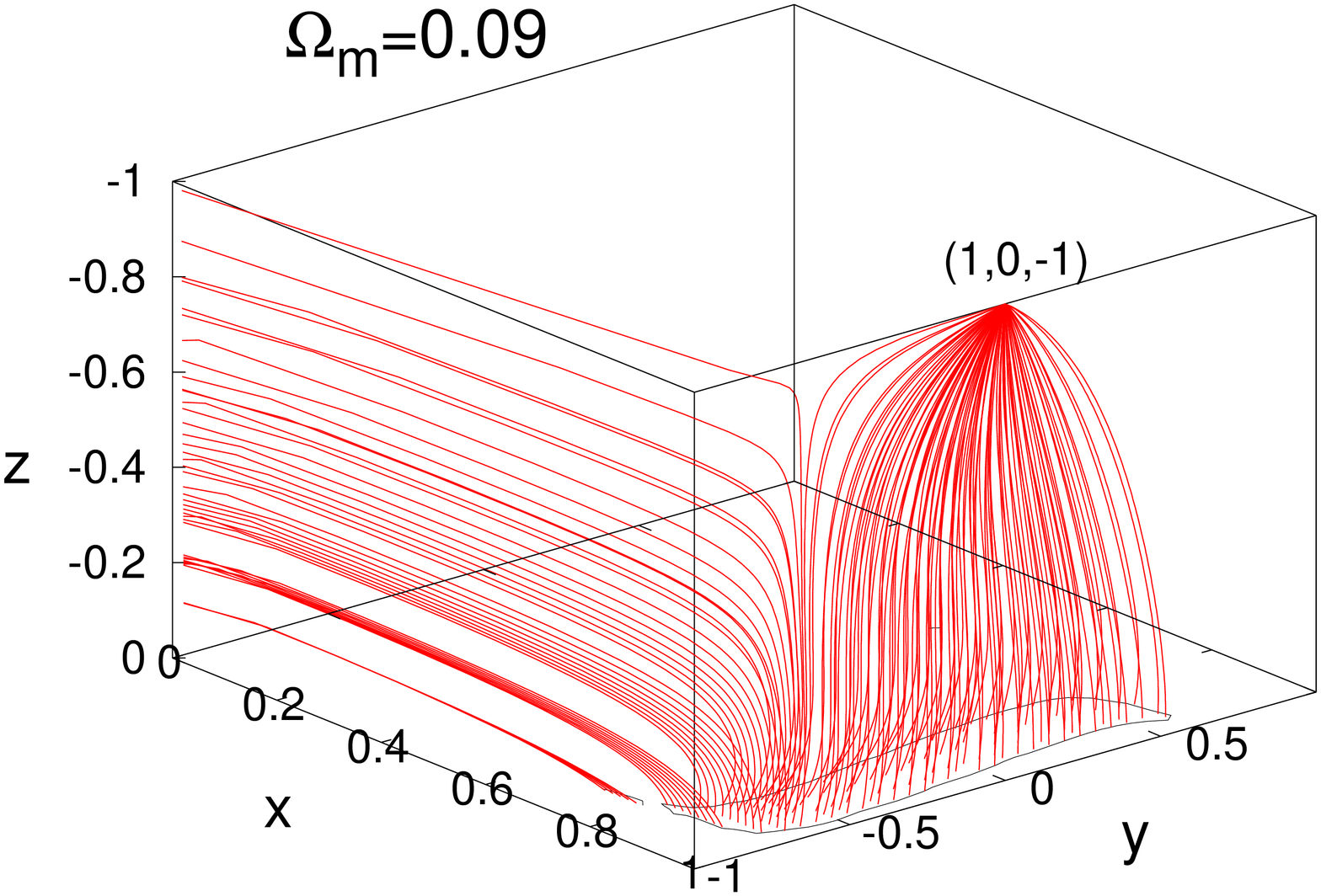}
\newline
\vskip 0.4cm
\includegraphics[height=8cm, angle=270]{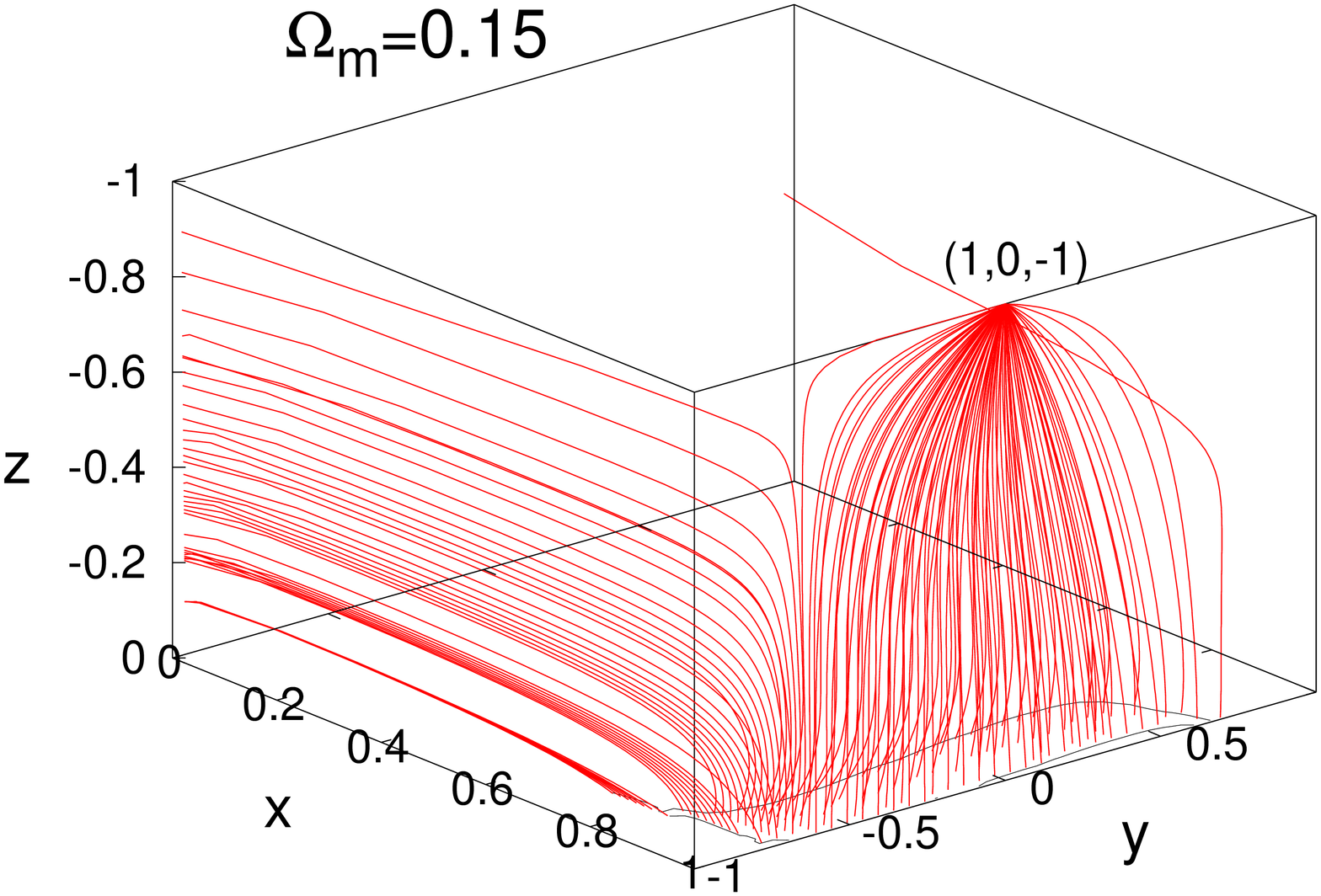}
\hskip 0.4cm
\includegraphics[height=8cm, angle=270]{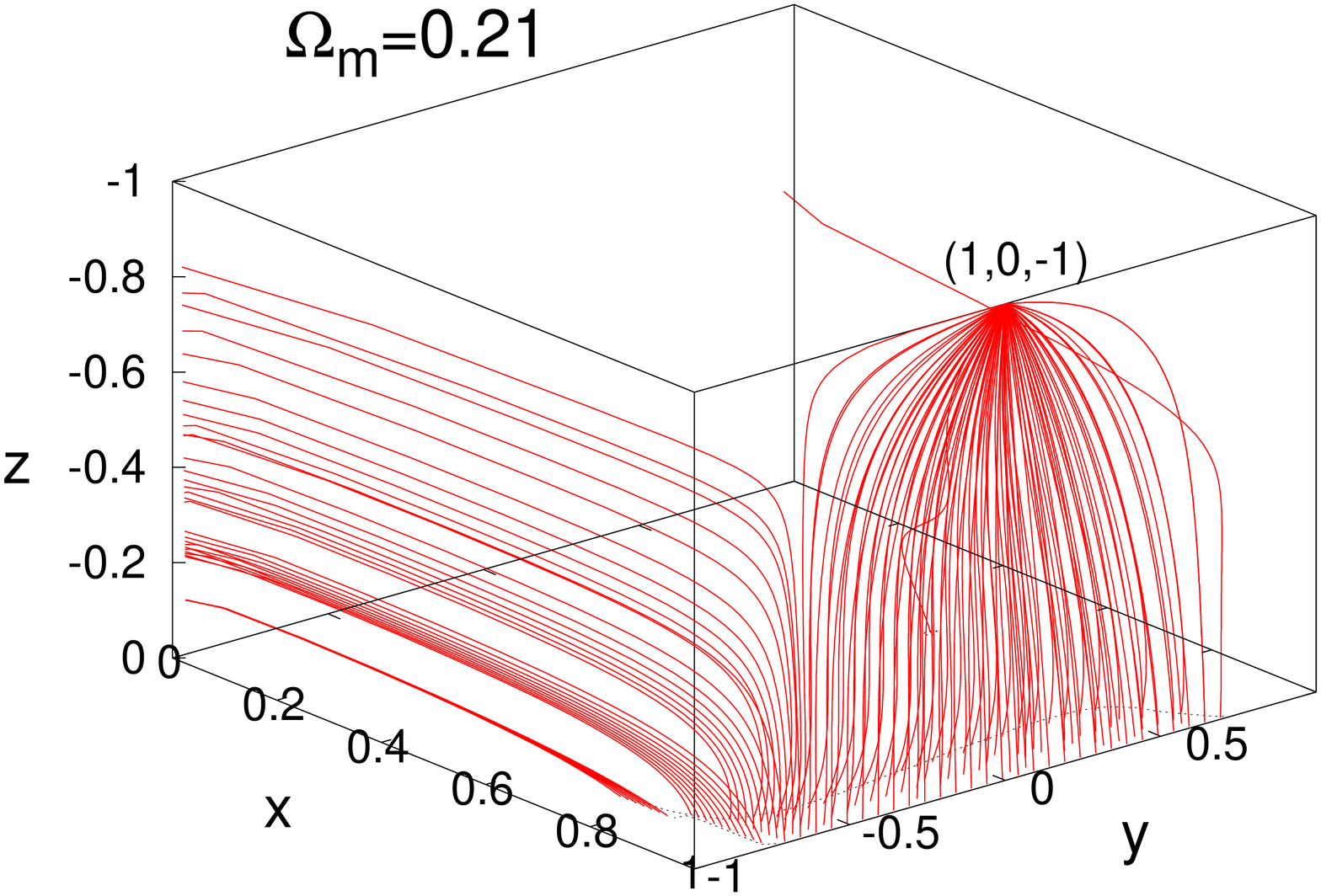}
\newline
\vskip 0.4cm
\includegraphics[height=8cm, angle=270]{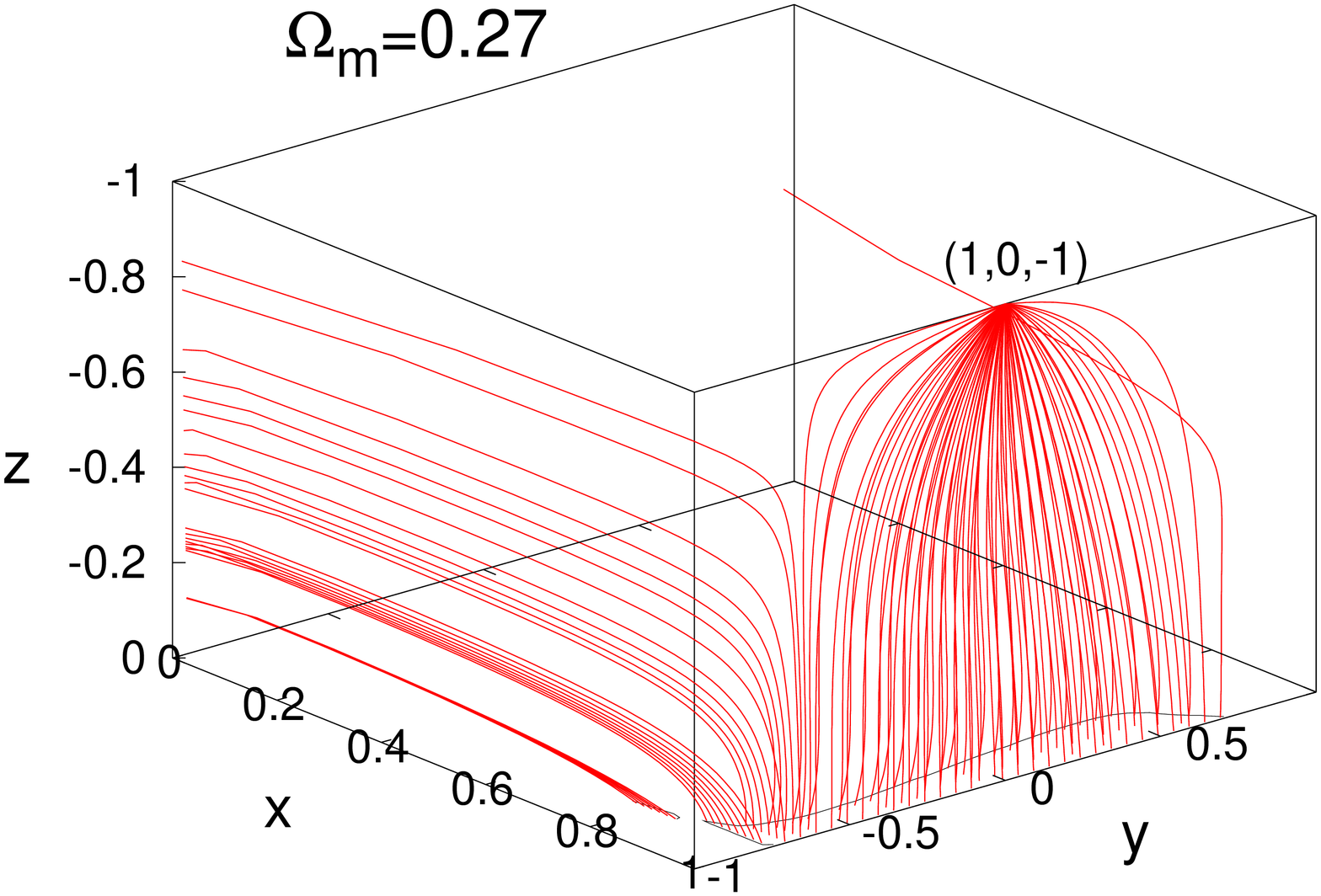}
\hskip 0.4cm
\includegraphics[height=8cm, angle=270]{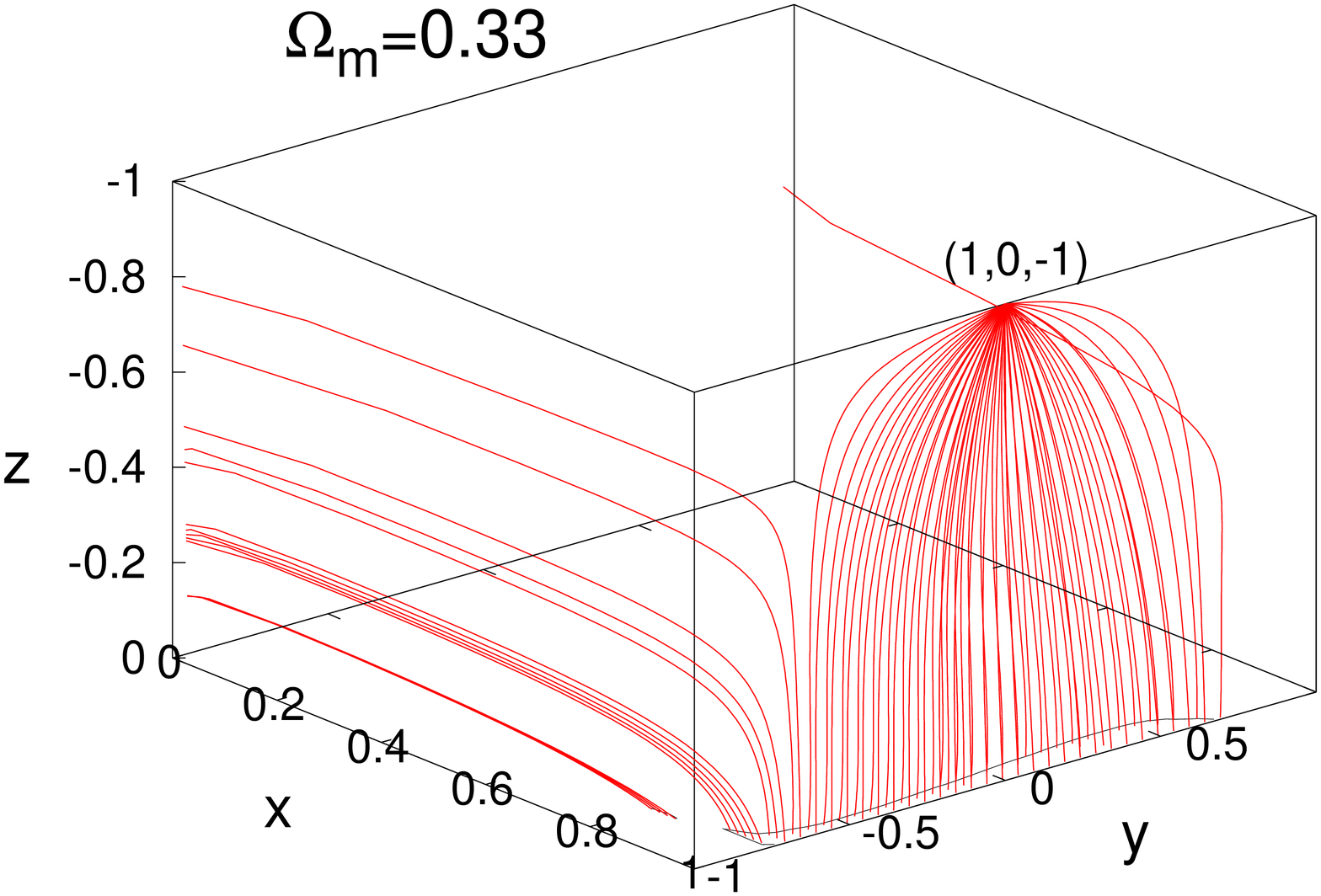}
\newline
\vskip 0.4cm
\caption{(Color online) The future evolution of those universes, which are
in a $68.3\%$ confidence level fit with the supernova data. The 1$\protect
\sigma $ contours (black lines in the $z=0$ plane) are from Fig \protect\ref
{Fig1} (the parameter plane $\left( y_{0},x_{0}\right) $ is the $z=0$ plane
here). The sequence of figures and the values of $\Omega_{m,0}$ are the same 
as on Fig. \protect\ref{Fig1}. 
The point  $(1,0,-1)$ is the de Sitter final state.}
\label{Fig3}
\end{figure}

\begin{table}[t]
\caption{
Key times in the evolution of tachyon universes for $k=0.4$ and $\Omega _{m,0}=0.03$ 
are given. 
The first two columns give initial values $x_{0}$, $y_{0}$ in agreement with 
supernovae data at the 1$\protect\sigma $ confidence level. Remaining columns 
starting from left give the successive times (measured from the present time): 
$t_{dec}$ when the expansion becomes 
decelerated, the tachyonic tansition time $t_{\ast}$, the first soft singularity 
crossing time $t_{S_{1}}$, the turning point $t_{turn}$, the second soft singularity 
crossing time $t_{S_{2}}$ and finally the Big Crunch time $t_{BC}$.
Times are given in $10^{9}yrs$ unit and  calculated
assuming $H_{0}=70$ km s$^{-1}$ Mpc$^{-1}$.}
\begin{center}
\begin{tabular}{c|c|c|c|c|c|c|c}
$y_{0}$ & $x_{0}$ & $t_{dec}$ & $t_{\ast }$ & $t_{S_{1}}$ & $t_{turn}$ & $%
t_{S_{2}}$ & $t_{BC}$ \\ \hline
$-0.80$ & $0.725$ & $0.60658$ & $0.89098$ & $1.71951$ & $1.71980$ & $1.72032$
& $2.01708$ \\ 
$-0.80$ & $0.755$ & $0.65334$ & $0.90487$ & $1.69670$ & $1.69698$ & $1.69747$
& $1.99019$ \\ 
$-0.75$ & $0.875$ & $2.24655$ & $2.48989$ & $3.29101$ & $3.29119$ & $3.29153$
& $3.59233$ \\ 
$-0.65$ & $0.875$ & $6.64484$ & $6.87894$ & $7.69232$ & $7.69237$ & $7.69247$
& $8.00317$ \\ 
$-0.60$ & $0.845$ & $9.29396$ & $9.52553$ & $10.34418$ & $10.34420$ & $%
10.34425$ & $10.65883$%
\end{tabular}%
\end{center}
\label{Table1}
\end{table}

\begin{table}[t]
\caption{Same as in Table \protect\ref{Table1} for $\Omega_{m,0}=0.27$.}
\begin{center}
\begin{tabular}{c|c|c|c|c|c|c|c}
$y_{0}$ & $x_{0}$ & $t_{dec}$ & $t_{\ast }$ & $t_{S_{1}}$ & $t_{turn}$ & $%
t_{S_{2}}$ & $t_{BC}$ \\ \hline
$-0.80$ & $0.770$ & $0.65390$ & $1.05684$ & $1.90378$ & $1.91751$ & $1.93727$
& $2.24682$ \\ 
$-0.75$ & $0.875$ & $2.53938$ & $2.90537$ & $3.78730$ & $3.79630$ & $3.80968$
& $4.13589$ \\ 
$-0.75$ & $0.950$ & $3.11633$ & $3.43361$ & $4.28179$ & $4.28892$ & $4.29969$
& $4.62210$ \\ 
$-0.70$ & $0.965$ & $6.64092$ & $6.92923$ & $7.81031$ & $7.81302$ & $7.81738$
& $8.15389$ \\ 
$-0.65$ & $0.995$ & $21.07994$ & $21.33077$ & $22.22677$ & $22.22681$ & $%
22.22689$ & $22.57314$%
\end{tabular}%
\end{center}
\label{Table2}
\end{table}

\end{widetext}

Now we can turn to the analysis of the Tables I and II. In Table \ref{Table1}
different times measured from today are given for a low amount of dust $%
\Omega_{m,0} = 0.03$ and in Table \ref{Table2} for $\Omega_{m,0} = 0.27$. So
we see the effect of the addition of dust in a systematic way. Three
comments are in order here.

First, the time interval $t_{\ast }$ between today and the first transition
into the pseudotachyon varies considerably within the set of trajectories
compatible with the supernovae data. Namely, it varies from 0.9 to 9.5
billion years for $\Omega _{m,0}=0.03$ and from 1 to 21 billion years $%
\Omega _{m,0}=0.27$. Second, the time intervals between $t_{\ast }$ and the
Big Crunch time $t_{BC}$ are practically constant (about 1.1 billion years
for the first case and about 1.2 billion years for the second case). A
similar property was found in the model without dust \cite{tach2}. Third,
the time interval between the two soft singularity crossings $%
t_{S_{2}}-t_{S_{1}}$ decreases strongly (from 810 thousand years to 70
thousand years for the first case and from 0.03 to 0.0002 billion years for
the second case) when the value of $t_{\ast }$ increases. This can be
ascribed to the fact that the density of dust, at the moment of the first
soft singularity crossing $t_{S_{1}}$, for the universes with high values of 
$t_{\ast }$ is greatly reduced compared to those with small values of $%
t_{\ast }$. Indeed, in the absence of dust the two values $t_{S_{1}}$ and $%
t_{S_{2}}$ coincide and we have a unique Big Brake singularity.

On Figure 2 the evolutions are shown in the three-dimensional coordinate
space $x,y,z$ for six different values of $\Omega_{m,0}$. For the
trajectories ending in a de Sitter space, the final point has coordinates $%
(1,0,-1)$. For other trajectories we present only the evolutions until the
first soft singularity crossing. Generally, the sets of initial conditions,
compatible with the supernovae data (the regions in the plane $(x,y)$ at $z
= 0$) decrease as the quantity of dust increases and vanish for $%
\Omega_{m,0} > 0.33$. Also, as $\Omega_{m,0}$ increases, the number of
trajectories going to a soft singularity is decreasing compared to those
ending in a de Sitter space.

This work is done with the spatially-flat paradigm in mind. However, as this
model is constrained using SNIa data, it is interesting to relax the
assumption of flatness in this case and to consider also non-flat universes
with a spatial curvature allowed by observations. Indeed in a spatially
closed universe the curvature and matter terms in the Friedmann equation
could cancel each other at some (negative) redshift $z_{curv}$ 
\begin{equation}
1+z_{curv} = \frac{|\Omega_{k,0}|}{\Omega_{m,0}}~.  \label{zcurv}
\end{equation}
The quantity $\Omega_{k,0}$ is strongly constrained by observations, $%
-0.0065 \le \Omega_{k,0} \le 0.0012$ (95\% C.L.) with central value $%
\Omega_{k,0}=-0.0027$ \cite{WMAP12}. Hence a slightly spatially closed
universe is favoured. 

Of course the tachyon, like any scalar field model and in sharp contrast to
the anti-Chaplygin gas, does not have a barotropic equation of state.
Therefore the amount of expansion needed to reach the soft singularity
depends on the initial conditions. It is quite clear however that for models
studied here we will have $|z_{S_1}|<|z_{curv}|$. Hence the kind of problem
considered in this paper, and the mechanism suggested in order to cross the
soft singularity, will remain even in the presence of a tiny curvature. But
we conjecture that peculiar initial conditions do exist for which this is no
longer the case.

\section{Concluding Remarks}

Soft cosmological singularities known since the 1980s \cite{Tipler2}, have
been attracting growing attention during the last few years \cite{soft}. In
this paper we have continued the investigation of particular cosmological
models based on tachyon fields or perfect fluids (introduced in paper \cite%
{tach0}), for which soft singularities arise in a natural way. The main
result of our investigation is the description of a smooth crossing of soft
singularities, arising in models with anti-Chaplygin gas or of a particular
tachyon field in the presence of dust. Such a crossing is accompanied by
certain transformations of matter properties, embodied in a change either of
equation of state or of Lagrangian.

The interesting feature of the tachyon model is that there exist
cosmological evolutions whose past is compatible with the supernova data and
whose future reveals \textquotedblleft exotic phase
transitions\textquotedblright\ which are described here in detail. We have
performed a detailed numerical analysis of these evolutions. 

All our studies, both theoretical and numerical, were performed assuming a a
spatially-flat universe. Next interesting step for the study of dark energy
models possessing soft future singularities is the inclusion of spatially
closed universes. Indeed, observations do allow for a tiny spatial
curvature, a positive curvature being slightly preferred. While a tiny
viable curvature will not change the situation for most models studied in
this paper, a larger number of situations can arise in the presence of
spatial curvature for the tachyon models because of their rich dynamics.
Indeed, if the universe reaches the point of maximal expansion before
occurence of the soft future singularity, the latter will not occur at all.
In the case of our tachyon model this can happen for specific initial
conditions. If for some peculiar initial conditions the turning point and
the soft singularity coincide the latter retains its character of a Big
Brake singularity. (In another dark energy model, based on a standard scalar
field, such an interplay between turning point and the encounter with a soft
singularity was considered in \cite{manti}). For a comprehensive
investigation of these situations a more detailed study is required, both
theoretical and numerical and this is left for future work \cite{future}. In
contrast, the possible situations in the case of the anti-Chaplygin gas are
more straightforward.

Another interesting direction of development of the present work is the
consideration of cosmological perturbations and their possible influence on
the structure of sudden singularities and on the conditions of their
crossing. 
To our knowledge no systematic study of this kind appeared yet in the
literature.

\section*{ACKNOWLEDGMENTS}

We thank the referee for the clarifying comments which led to the inclusion
of subsection III.D. The work of Z.K. was supported by OTKA\ Grant No.
100216, L.\'{A}.G. was supported by the European Union / European Social
Fund Grant No. T\'{A}MOP-4.2.2.A-11/1/KONV-2012- 0060, and A.K. was
partially supported by the RFBR Grant No. 11-02-00643.

\end{document}